# Hidden Warming Effects of Cloud Cycle Errors in Climate Models

Jun Yin[1], Amilcare Porporato[1,2]*

**Affiliations:**

[1]Department of Civil and Environmental Engineering, Duke University, Durham, North Carolina, USA.
[2]Nicholas School of the Environment, Duke University, Durham, North Carolina, USA.
*Correspondence to: amilcare.porporato@duke.edu

**Abstract:**

Clouds' efficiency at reflecting solar radiation and trapping the terrestrial one is strongly modulated by their diurnal cycle. Much attention has been paid to mean cloud properties due to their critical role in climate projections; however, less research has been devoted to their diurnal cycle. Here, we quantify the mean, amplitude, and phase of the cloud cycles in current climate models and compare them with satellite observations and reanalysis data. While the mean values appear to be reliable, the amplitude and phase of the diurnal cycles of clouds show marked inconsistencies, especially over land. We show that, to compensate for the increased net radiation input implied by such errors, an overestimation of the cloud liquid water path may be introduced during calibration of climate models to keep temperatures in line with observations. These discrepancies are likely to be related to cloud parametrization problems in relation to atmospheric convection.

**One Sentence Summary:**

Errors in diurnal cycle of clouds induce warming effects in climate models, likely compensated by the overestimation of cloud liquid water path.

**Main Text:**

As efficient modulators of the Earth's radiative budget, clouds play a crucial role in making our planet habitable (*1*). Their response to the increase in anthropogenic emissions of greenhouse gases will also have a substantial effect on future climates, although it is highly uncertain whether this will contribute to intensifying or alleviating the global warming threat (*2*). Such uncertainties are well recognized in the state-of-the-art General Circulation Models (GCMs) (*3*) and are typically associated with their performance in simulating some critical cloud features, such as cloud structure and coverage (*4*). Among these features, perhaps the most overlooked one is the diurnal cycle of clouds (DCC), describing how certain cloud properties (e.g. cloud coverage) change throughout the day at a given location. Due to the clouds' interference with the diurnal fluctuations of solar and terrestrial radiation, shifts in DCC have the potential to strongly affect the Earth's energy budget, even when on average the daily cloud coverage is the same (*5*). It is thus clear that a reliable representation of DCC in GCMs is fundamental for climate projections (*6*).

To assess the degree with which climate models capture the key features of the DCC, we calculate three main statistics describing the typical DCC in each season in climate model outputs and compare them with those obtained from satellite observations and reanalysis data. The potential impacts of the resulting discrepancies on the radiative balance and global temperatures are then assessed using a radiative balance model. Specifically, we consider the cloud coverage, whose diurnal cycle is closely related to that of total cloud water path (*7*, *8*) and thus plays a critical role in the energy budget. To avoid dealing with higher harmonics of a Fourier decomposition of the DCC for cases with significant deviations from sinusoidal shapes (*9*) and lack of diurnal periodicity, here we focus on the standard deviation ($\sigma$), centroid ($c$), and mean ($\mu$) to capture the amplitude, phase, and the daily average of cloud coverage (see 1st column in Fig. 1 and Methods). These three indexes of cloud climatology are computed for the outputs of the GCMs participating in the Fifth Phase of the Coupled Model Inter-comparison Project (listed in Table S1), and then are compared with those from the International Satellite Cloud Climatology Project (ISCCP) (*10*) and from the European Centre for Medium-Range Weather Forecasts (ECMWF) twentieth century reanalysis (ERA-20C) (*11*), all of which have high-frequency (3-hour) global coverage for the period 1986-2005. While we are well aware that the ISCCP satellite records contain artifacts which may affect long-term trends, it is important to emphasize here that they do provide very useful information about the cloud climatology (*12*, *13*) of interest here. Regarding ERA-20C, it is the ECMWF's state-of-the-art reanalysis designed for climate applications (*11*). Note that both ERA-20C reanalysis and CNRM-CM5 climate model rely on the same Integrated Forecast System from ECMWF (*11*, *14*), so that some common elements of cloud climatology may be expected.

The first column in Fig. 1 shows an example of DCC climatology and the corresponding indexes for a subtropical monsoon climate zone in Eastern China in summer, characterized by clear mornings and frequent afternoon thunderstorms. This type of diurnal cycle is evident from the satellite (1st row) and reanalysis data (2nd row) but is not captured by the GCMs (3rd row). Similar discrepancies are also found in other regions across the various GCMs (see Fig S1). To explore these discrepancies globally, we calculate the DCC indexes at each grid point in each season from each data source (shown in Fig S6-Fig S15). The most striking feature of DCC indexes are the land/ocean patterns, reflecting the contrasting mechanisms of atmospheric convections, although these geographical patterns are less coherent in the GCM outputs. For this reason, we compare the empirical distributions of DCC indexes over the land and ocean in the 2nd-4th column of Fig. 1 (and in Fig S2). The satellite and reanalysis data clearly show larger $\mu$, smaller $\sigma$, and earlier $c$ over the ocean. A consistent pattern is found in GCM outputs for the mean cloud coverage (2nd column in Fig. 1). However, the DCC amplitude $\sigma$ generally shows no significant land-ocean contrast and a number of GCMs erroneously suggest stronger DCC amplitude over the oceans (3rd column), while regarding the phase $c$, the land-ocean contrast is underestimated with most GCMs not even capturing the afternoon cloud peaks (4th column).

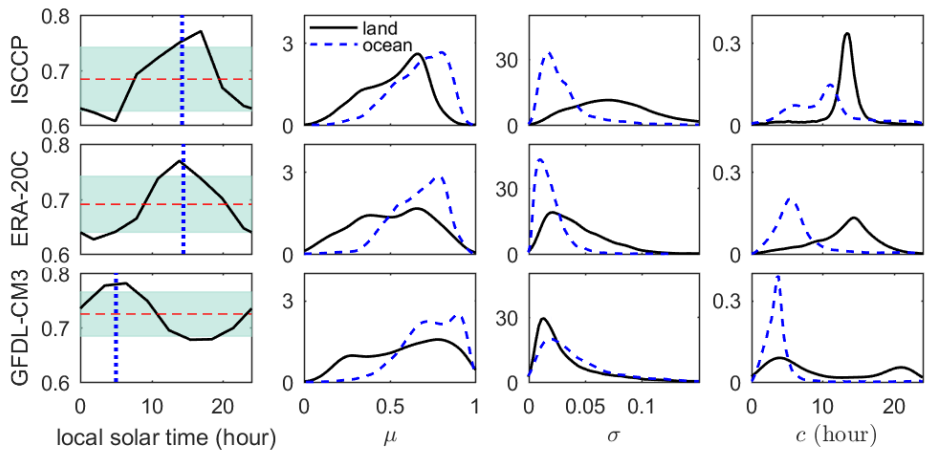

**Fig. 1.** DCC climatology and its indexes from ISCCP satellite records (1st row), and ERA-20C reanalysis data (2nd row), and GFDL-CM3 climate model 'historical' experiment (3rd row). More results from multiple GCM outputs are presented in Fig S1 and Fig S2. (**1st column**) Examples of diurnal cycle of average cloud coverage near Guangde, Anhui, China (30.7N, 119.2E) in summer (June, July, and August) averaged over 1986-2005. The vertical dot lines and horizontal dash lines show the centroid and mean of the diurnal cycle climatology; the shaded blue areas indicate plus and minus one standard deviation. (**2nd-4th column**) Empirical probability density function (PDF) of mean ($\mu$), standard deviation ($\sigma$), and centroid ($c$) of diurnal cycle of cloud coverage climatology over the land (black solid lines) and ocean (blue dash lines) in all four seasons over 60S-60N.

A detailed analysis of these differences is given in Fig. 2, which compares the root-mean-square deviation (RMSD, see Methods) of $\mu$, $\sigma$, and $c$ among climate models, reanalysis data, and satellite observations over the land (upper triangle matrix) and ocean (lower triangle matrix). In general, models with obvious similarities in code produce a similar cloud climatology and thus have smaller RMSD (e.g. CNRM-CM5 and ERA-20C; GFDL-ESM2M and GFDL-ESM2G). More specifically, for $\mu$, the values are relatively homogeneous with a certain degree of symmetry, showing similar land/ocean differences between each pair of models. The corresponding Taylor diagrams (*15*) further suggest that the mean cloud coverage is much better simulated than the rest DCC indexes (see Fig S4 and Fig S5). For $\sigma$, the dominant RMSD between ISCCP records and the other datasets in the upper matrix indicates that all climate models fail to capture the strong amplitude of DCC over land. Finally, regarding $c$, the much larger RMSD in the upper matrix suggests that over the continental regions the DCC phases in GCMs not only deviate from the observations but also have considerable inter-model differences.

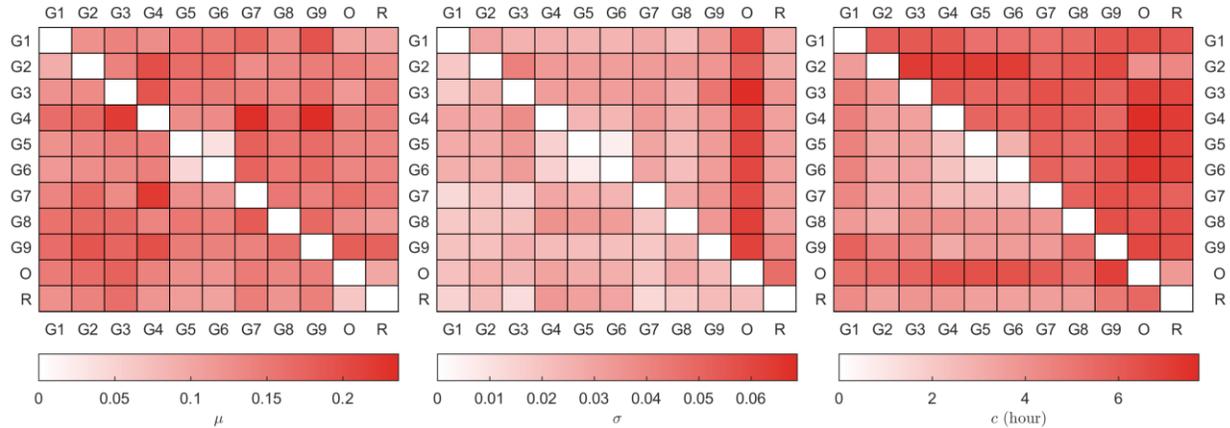

**Fig. 2.** Root-mean-square deviation (RMSD) of mean ($\mu$), standard deviation ($\sigma$), and centroid ($c$) of diurnal cycle of cloud coverage from climate models (G1-G9, full names are listed in Table S1), ISCCP satellite observations (O), and ERA-20C reanalysis data (R). Each element in the color matrix is corresponding to the RMSD from two data sources marked by the abbreviations on the axes. The lower and upper triangle matrixes are for the RMSD over the ocean and land, respectively, and the diagonal matrix, RMSD in itself, is zero by definition.

Given the discrepancies just discussed, it is logical to wonder about their potential impacts in climate predictions. Although providing a precise answer to this difficult question would require an in-depth analysis of the detailed spatial and temporal structures of cloud climatology, for the purposes of this investigation, a reasonably robust way to gain conceptual understanding of the DCC controls on the earth energy balance may be offered by global radiation balance considerations (see Methods), following previous analysis of global greenhouse gas effects and water vapor feedbacks (*16*, *17*). In this approach, the troposphere is assumed to be constituted of clear air and clouds with variable fractional coverage, *f*(*t*). The diurnal cycle of solar radiation is approximated by a parabolic function and a finite heat capacity is assigned to the earth surface layer to model its temperature variation at the diurnal time scale. Varying the amplitude and phase of the cloud fraction, while keeping the mean cloud coverage and other parameters constant, we can analyze the changes in a hypothetical surface temperature in one diurnal cycle as a function of $c$ and $c_v = \sigma / \mu$ (see Methods). This mean surface temperature, hereafter referred to as reference temperature, is visualized as a 'heatmap' in Fig. 3 as a function of the DCC indices. The reference temperature is symmetric with respect to the centroid at noon ($c =$ 12 hr) with colder temperature for midday cloud peak. As one would expect, clouds only have warming effects at night, such that earlier cloud phases (i.e., before sunrise) inevitably induce warming regardless of the cloud types and structures; similarly, clouds always have cooling effects when the solar radiation is strong during the day so that midday cloud peaks typically induce cooling. Such impacts of phase ($c$) become more significant under larger relative amplitude ($c_v$). For example, for $c_v$=0.1, the reference temperature increases by 9.3 K in response to a shift of the centroid from noon to midnight, while for $c_v$=0.2, the increase of temperature becomes 18.7 K for the same centroid shift. This large temperature change is consistent with the values reported in a prior study (*5*) and is due to the significant and systematic variations in radiation.

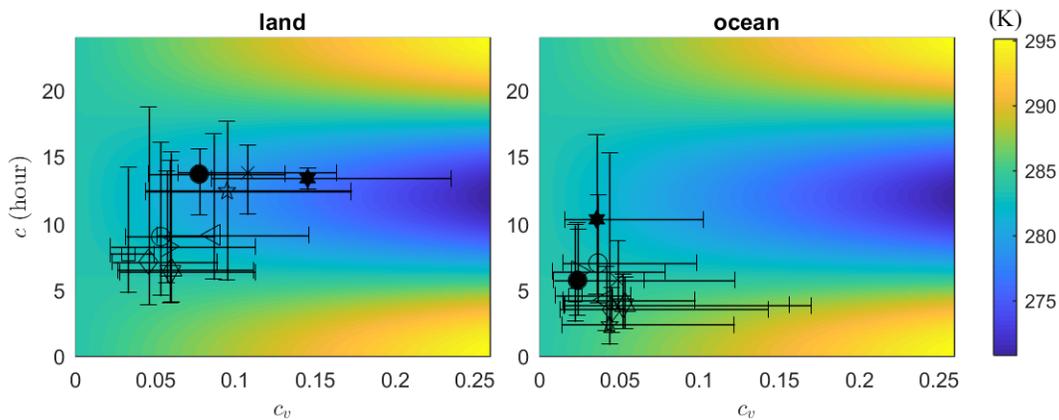

**Fig. 3.** 'Heatmap' of reference temperature as functions of coefficient of variation ($c_v$) and centroid ($c$) of cloud diurnal cycles. The crosses specify the 25th, 50th, and 75th percentiles of the $c$ and $c_v$ from GCMs (open markers), ISCCP records (filled hexagram), and ERA-20C reanalysis data (filled circle). Detailed symbols and their corresponding data sources are listed in Table S1.

To gain further insight, it is now easy to superimpose the $c$ and $c_v$ derived from GCMs, ISCCP, and ERA-20C onto the 'heatmap' obtained from the simplified radiation model. Over the land, the indexes appear much more scattered due to larger discrepancies of both $c$ and $c_v$ among each data source, as already illustrated in Fig. 2. The continental clouds tend to peak in the afternoon as observed in ISCCP and simulated in ERA-20C, reflecting more solar radiation and pushing the simulated climate system into the cold zone of the 'heatmap' (Fig. 3A). Over the ocean, the indexes are much more clustered due to the greater consistency between the DCC observations and simulations. Nonetheless, GCMs still seem to simulate a little earlier cloud peak, trapping more longwave radiation during the night and pushing the simulated climate system toward the warm zone (Fig. 3B).

Finally, we calculated the distributions of reference temperature using the joint distribution of $c$ and $c_v$ from each data source (see Fig. 4A). The results clearly show that the reference temperatures calculated with the DCC properties of the GCMs are systematically higher than those of ISCCP and ERA-20C. How is it then that GCM results give similar climatology of earth surface temperatures? While of course different explanations could be put forward for this, a likely reason is that the extra longwave radiation trapped over the oceans and the increased solar radiation absorbed on land may be balanced in GCMs by larger values in the mean liquid water path. This increased reflection of solar radiation would then compensate for the increased temperatures that the DCC errors would otherwise induce. To provide support to this hypothesis, Fig. 4B shows the relationship between the mean reference temperature and the cloud water path. As can be seen, the hotter the system is, the larger the cloud water path becomes, thus suggesting an artificial cloud thickening induced by temperature calibration. This is also in line with prior studies that showed how most climate models tend to overestimate the liquid clouds (*18*, *19*), which are more efficient at reflecting solar radiations than the ice clouds (*20*, *21*). As a result, while the radiation budgets in GCMs is correct on the daily averages and reproduces the observed surface mean temperature climatology, it remains inaccurate at the sub-daily scale.

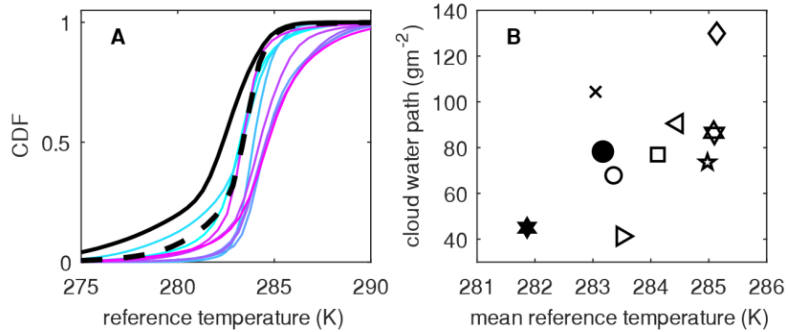

**Fig. 4.** (**A**) empirical cumulative density functions (CDF) of the reference temperature at the corresponding $c$ and $c_v$ within 60S-60N derived from the GCM outputs (cyan to magenta lines represent G1 to G9 as listed in Table S1), ISCCP records (black solid lines), ERA-20C reanalysis (black dash lines). (**B**) Relationship between global mean cloud water path and mean reference temperature from GCMs (open markers), ISCCP records (filled hexagram), and ERA-20C reanalysis data (filled circle). Detailed symbols and their corresponding data sources are listed in Table S1. The positive correlation coefficient between these two variables are statistically significant at the 0.05 level (one-tailed test).

In summary, we quantified the discrepancies of the diurnal cycle of clouds among current climate models, satellite observations, and reanalysis data. In general, climate models have better and more consistent performance in simulating mean cloud coverage, while they present considerable discrepancies in the standard deviation ($\sigma$) and centroid ($c$) of cloud cycles. The most problematic errors appear to be due to the smaller $\sigma$ and earlier $c$ over the land, leading to an overestimation of net radiation as indicated by a simple radiation model. In turn, it likely that to counteract this effect, the calibration of climate models actually causes an overestimation of the liquid water path (*18*, *19*). While this compensates for the warming effects due to errors in $\sigma$ and $c$, the effects of these errors in the cloud representation at the sub-daily timescale remain problematic and call for a deeper understanding of cloud dynamics and atmospheric convections to improve climate projections.

**Acknowledgments**

We acknowledge support from the USDA Agricultural Research Service cooperative agreement 58-6408-3-027; and National Science Foundation (NSF) grants EAR-1331846, EAR-1316258, FESD EAR-1338694 and the Duke WISeNet Grant DGE-1068871. The ISCCP satellite records were obtained from NASA Atmospheric Science Data Center (http://isccp.giss.nasa.gov/). The ERA-20C reanalysis data were obtained from European Centre for Medium-Range Weather Forecasts (http://www.ecmwf.int). The climate model data were downloaded from the fifth phase of the Coupled Model Intercomparison Project website (http://cmip-pcmdi.llnl.gov). Models used in the paper are available upon request.

**Supplementary Materials:**

Materials and Methods
Figures S1-S15
Tables S1
References (*22-33*)

# Supplementary Materials for

# Hidden Warming Effects of Cloud Cycle Errors in Climate Models

Jun Yin, Amilcare Porporato*

correspondence to: amilcare.porporato@duke.edu

**This PDF file includes:**

Materials and Methods
Supplementary Text
Figs. S1 to S15
Table S1

**Material and Methods**

Diurnal cycle of clouds

The time series of cloud coverage at each grid box (*i*) in each season (*j*) from each data source (*m*) were analyzed as follows. For the period 1986-2005, in each day the cloud coverage is given at three-hour interval (e.g., at local solar time $t_1$=3 hr, $t_2$=6 hr, … $t_k$=3·$k$ hr, …, $t_8$ =24 hr). We first average, by season, these series to obtain a typical diurnal cycle of cloud coverage,

$$\overline{f}_{mij}(t_1), \quad \overline{f}_{mij}(t_2), \quad ... \quad \overline{f}_{mij}(t_k), \quad ... \quad \overline{f}_{mij}(t_8), \tag{1}$$

where subscripts *m*, *i*, *j*, and *k* represent the data source index, grid location index, season index, and discrete time index, respectively. To characterize climatology of diurnal cycle of clouds (DCC), we define three indexes: the mean, amplitude, and phase.

The mean of the DCC is directly defined as the expectation

$$\mu_{mij} = \frac{1}{8}\sum_{k=1}^{8}\overline{f}_{mij}(t_k). \tag{2}$$

The amplitude of the DCC is quantified by its corrected sample standard deviation as

$$\sigma_{mij} = \sqrt{\frac{1}{7}\sum_{k=1}^{8}\left[\overline{f}_{mij}(t_k) - \mu_{mij}\right]^2}. \tag{3}$$

The coefficient of variation can be expressed as

$$\left(c_v\right)_{mij} = \frac{\sigma_{mij}}{\mu_{mij}}. \tag{4}$$

The latter is useful to analyze the impact of relative amplitude of DCC across different models.

The phase of the DCC is given by the centroid of $t_k$ weighted by the probability distribution of cloud coverage during one diurnal cycle

$$p_{mij}(t_k) = \frac{\overline{f}_{mij}(t_k)}{\sum_{k=1}^{8}\overline{f}_{mij}(t_k)}. \tag{5}$$

Since $t_k$ within one diurnal period can be treated as a circular quantity, the calculation of centroid ($c$) uses the circular statistics (*22*),

$$c_{mij} = \frac{\tau}{2\pi}\arg\left[\sum_{k=1}^{8}p_{mij}(t_k)\exp\left(\mathbf{i}\frac{2\pi t_k}{\tau}\right)\right]. \tag{6}$$

where **i** is the imaginary unit and $\arg[\cdot]$ is the argument of a complex number. As can be seen in the examples of Fig. 1 and Fig S1, the centroid is located around the timing of the most cloudiness in one typical day.

The root-mean-square deviation (RMSD) of $\mu$ between data source $m_1$ and $m_2$ is defined as,

$$R_\mu(m_1, m_2) = \sqrt{\frac{1}{JI} \sum_j \sum_i (\mu_{m_1 ij} - \mu_{m_2 ij})^2} ,  \tag{7}$$

where $I$ and $J$ are the numbers of grid boxes and seasons considered in the calculation of the corresponding RMSD. Similarly, the RMSD of $\sigma$ is,

$$R_\sigma(m_1, m_2) = \sqrt{\frac{1}{JI} \sum_j \sum_i (\sigma_{m_1 ij} - \sigma_{m_2 ij})^2} ,  \tag{8}$$

and the RMSD of $c$ is,

$$R_c(m_1, m_2) = \sqrt{\frac{1}{JI} \sum_j \sum_i \left(c_{m_1 ij} - c_{m_2 ij} + n\tau\right)^2} .  \tag{9}$$

where $n$ is an integer and $\tau$ is the length of one diurnal cycle (24 hours). The integer $n$ is properly chosen such that the centroid difference ($c_{m_1 ij} - c_{m_2 ij} + n\tau$) is within $[-\tau/2, \tau/2]$.

Minimalist Radiative Balance Model

To estimate the impacts of DCC on the earth surface temperature, we employ a minimalist radiative balance model (*16*, *17*), a schematic diagram of which is shown in Fig S3. The troposphere is assumed to be comprised of clear air and clouds, the latter of which characterized by a typical diurnal variation of cloud coverage $f(t)$. For simplicity we model $f(t)$ as

$$f(t) = \mu + \sqrt{2}\sigma \cos\left[w(t-c)\right].  \tag{10}$$

The total shortwave radiation absorbed by the earth surface is

$$\begin{aligned} R_s(t) &= [1-f(t)]S(t)(1-\alpha_s) + f(t)S(t)\left[(1-\alpha_c - a_c)(1-\alpha_s) + (1-\alpha_c - a_c)\alpha_s\alpha_c(1-\alpha_s) + ...\right] \\ &= [1-f(t)]S(t)(1-\alpha_s) + f(t)S(t)(1-\alpha_c - a_c)(1-\alpha_s)\frac{1}{1-\alpha_s\alpha_c}, \end{aligned} \tag{11}$$

where $\alpha_c$ and $a_c$ are cloud albedo and absorptivity, $\alpha_s$ is surface albedo, and $S$ is the solar radiation reaching the tropopause. Accordingly, the net longwave radiation at the earth surface is

$$R_l(t) = [1-f(t)]\left[\sigma\varepsilon_g T_g^4(t) - \sigma T_s^4(t)\right] + f(t)\left[\sigma\varepsilon_c T_c^4(t) - \sigma T_s^4(t)\right], \tag{12}$$

where $\sigma$ is Stefan–Boltzmann constant, $\varepsilon_g$ is bulk longwave emissivity of the simplified greenhouse gas layer in the clear sky, $\varepsilon_c$ is the bulk longwave emissivity of the cloudy atmosphere, and $T_c$, $T_g$, and $T_s$ are the temperature of the cloudy layer, the greenhouse gas layer, and the earth surface.

We assume equilibrium between the greenhouse gas layer and cloudy layer so that

$$2\sigma\varepsilon_g T_g^4(t) = \sigma\varepsilon_g T_s^4(t), \tag{13}$$

and

$$2\sigma\varepsilon_c T_c^4(t) = \sigma\varepsilon_c T_s^4(t) + S(t)a_c + S(t)(1-\alpha_c - a_c)\alpha_s a_c + ...$$
$$= \sigma\varepsilon_c T_s^4(t) + S(t)a_c + S(t)(1-\alpha_c - a_c)\frac{\alpha_s a_c}{1-\alpha_s \alpha_c}. \tag{14}$$

These equations connect $T_c$ and $T_g$ to $T_s$. The net radiation at the earth surface is the sum of the net shortwave and longwave radiations,

$$R(t) = R_s(t) + R_l(t). \tag{15}$$

To model the temperature fluctuation at diurnal scale, we consider the energy storage in the earth surface layer so that the heat storage can be expressed as

$$C\frac{dT_s(t)}{dt} = R(t), \tag{16}$$

where $C$ is the heat capacity of the earth surface layer.

We set $S(t)$ as a parabolic function with peak at noon and zero during night (6pm-6am). The mean value of $S(t)$ during the whole diurnal cycle is 1/4 of the solar constant, which is a ratio of earth's cross-section area to its surface area accounting for the earth's rotation effects (*16*). The surface albedo is set as 0.06 to represent the typical ocean surface. The surface layer heat capacities are set as $8.4\times 10^5$ and $8.4\times 10^6$ J K$^{-1}$ m$^{-2}$ so that the diurnal temperature ranges are about 11 K and 1 K, representing typical temperature range over the land and ocean (*23*, *24*). This heat capacity influences the diurnal amplitude of surface temperature but has limited impacts on its mean value. The mean cloud coverage is set as 0.62 and in-cloud water path is set as 66 gm$^{-2}$, which are the mean values within 60S-60N derived from ISCCP records. The cloud albedo and shortwave absorption are modeled as empirical functions of liquid water path and solar zenith angle (*25*). The emissivity for clouds is set as unity and for greenhouse gas it is set as 0.78 (*26*).

The ordinary differential equation (16) is then solved numerically. After a periodic steady state is reached, the daily mean surface temperature is computed as

$$\overline{T_s} = \frac{1}{\tau}\int_0^\tau T_s(t)dt, \tag{17}$$

where $\tau$=24 hours. This mean temperature, referred to as reference temperature in the main text, is a function of $c$ and $c_v$, and can be conveniently used as a hypothetical temperature to compare the DCC impacts (see Fig. 3).

**Supplementary Text**

The following figures and table provide complementary information for illustrating the diurnal cycle of clouds and for understanding the radiative balance model. Specifically:

- Fig S1, corresponding to the 1st column in Fig. 1, provides additional examples of diurnal cycle of cloud climatology from multiple GCM outputs in three more sites.
- Fig S2, corresponding to the 2nd-4th column in Fig. 1, provides additional probability distribution functions of DCC indexes calculated from multiple GCM outputs.
- Fig S3 illustrates the radiation components analyzed in the radiative balance model in the 'Material and Methods' section.
- Fig S4 and Fig S5 provide additional information for comparing the relative discrepancy among three matrixes, as separately shown in Fig. 2.
- Fig S6-Fig S15 show the geographic distributions of DCC indexes from climate models, satellite records, and reanalysis data.
- Table S1 lists the names/institutions/symbols/identification numbers of climate models, satellite records, and reanalysis data. The identification numbers in Fig. 2 and the symbols in Fig. 3 and Fig. 4 are used to represent the corresponding data sources.

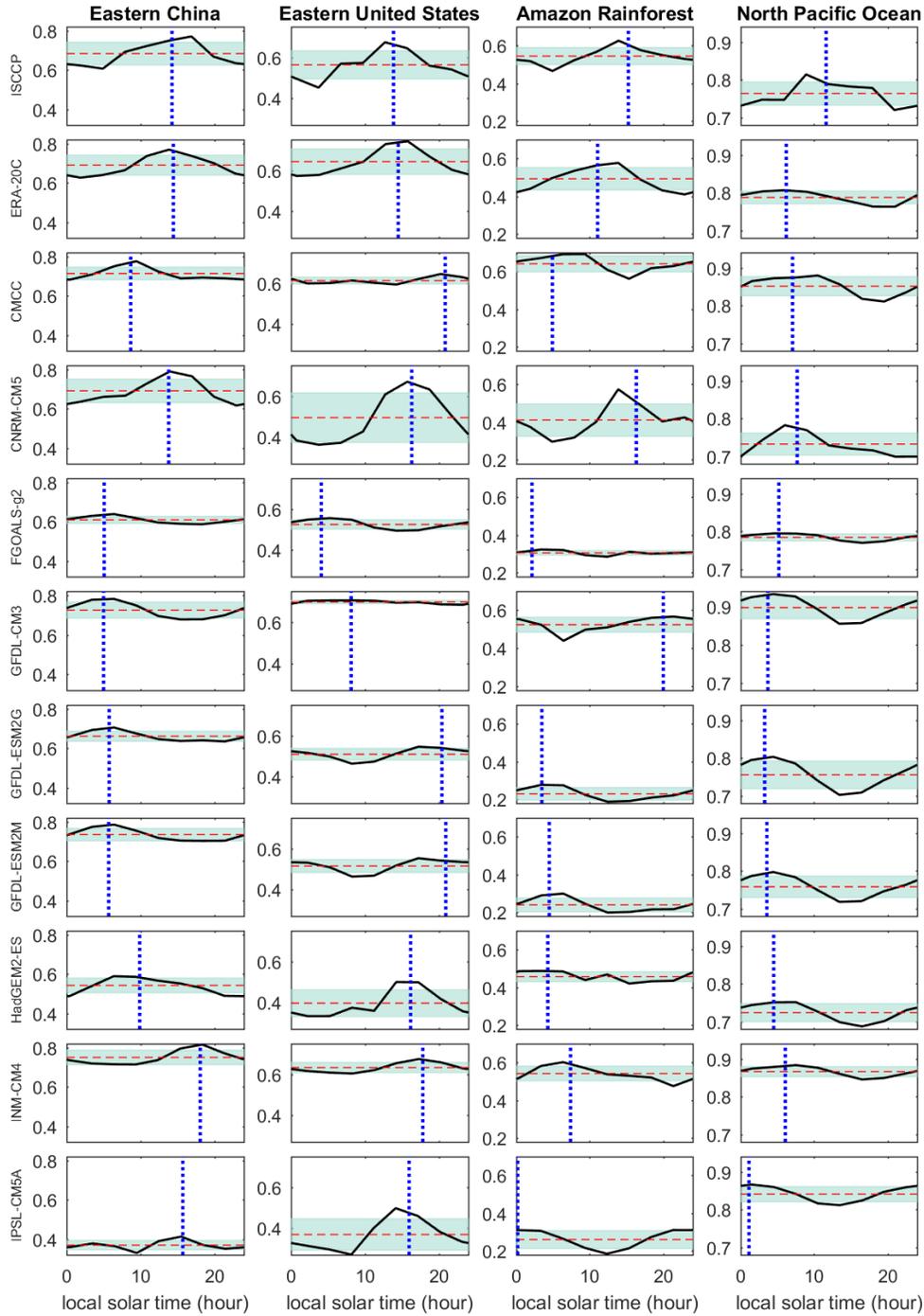

**Fig S1.** Examples of diurnal cycle of average cloud coverage in summer (June, July, and August) during 1986-2005 from ISCCP records (1th row), ERA-20C reanalysis (2nd row), and GCMs 'historical' experiment (3rd-11th row), near four locations (1st-4th column): Guangde, China (30.7N, 119.2E), Durham, USA (36N,79.9W), Codajás, Brazil (3.5S, 62.2W), and North Pacific Ocean (36S, 180E). The vertical dot lines and horizontal dash lines show the centroid and mean of the diurnal cycle climatology; the shaded blue areas indicate plus and minus one standard deviation.

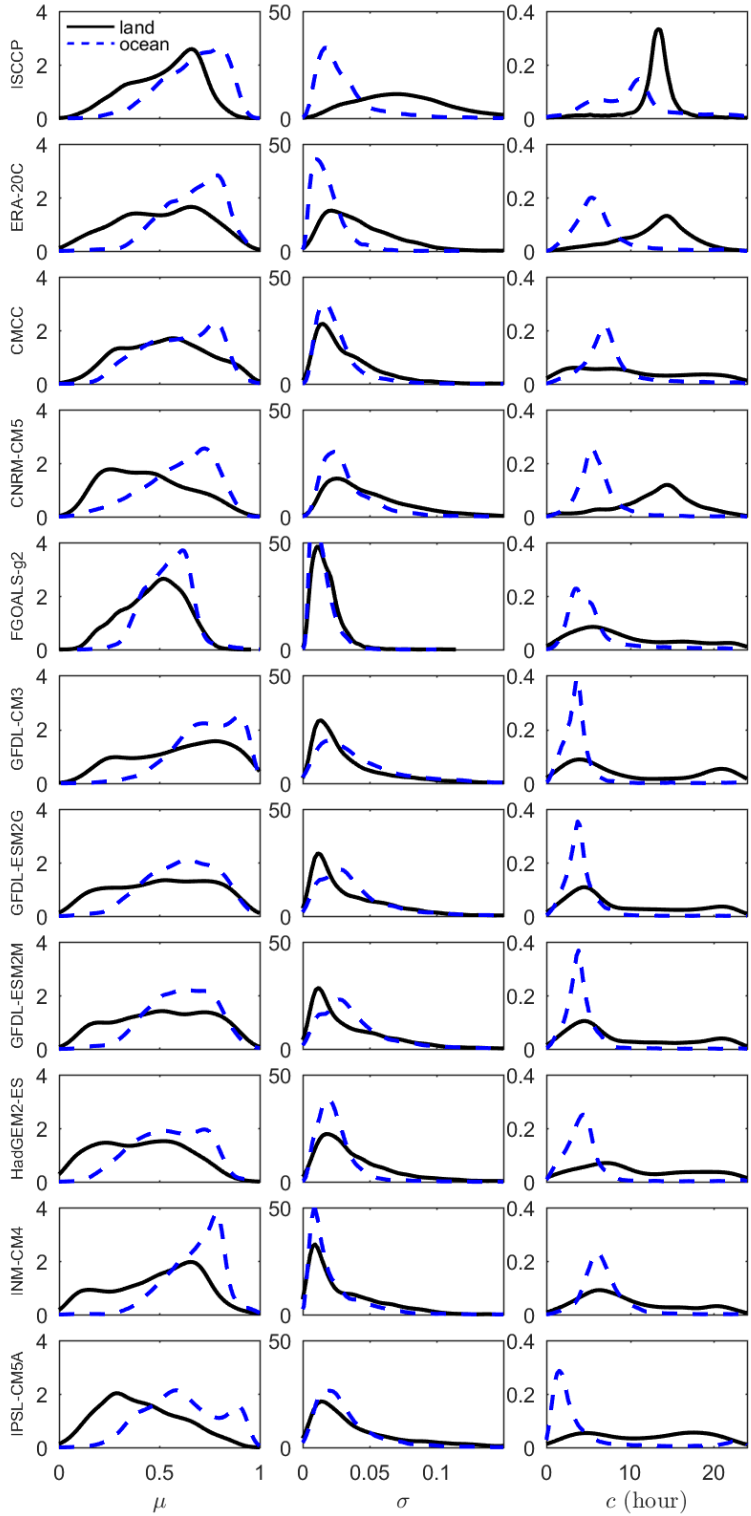

**Fig S2.** Empirical probability distribution of diurnal cycle climatology index from ISCCP records (1st row), ERA-20C reanalysis (2nd row), and nine GCM outputs (3rd-11th row) during 1986-2005 in all four seasons.

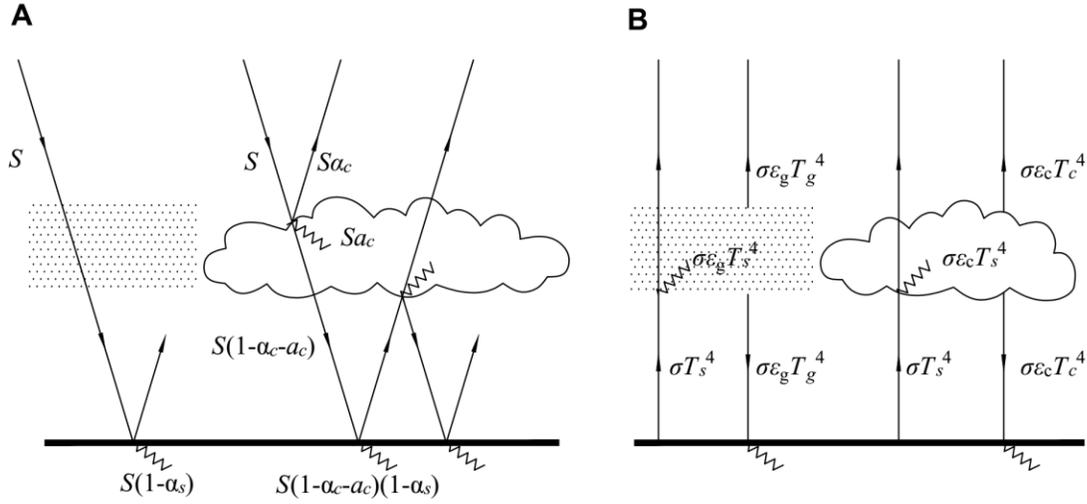

**Fig S3.** Schematic diagram of radiation components for the radiation balance model. The dot area and cloud-shaped area represent layers of clear and cloudy atmosphere, respectively. In (**A**), the solar radiation passes through the clear atmosphere, while it is partially reflected and absorbed by the clouds with albedo $\alpha_c$ and absorptivity $a_c$. In (**B**), the clear atmosphere is assumed to be a grey body with emissivity $\varepsilon_g$, which absorbs and reemits the longwave radiation; the clouds also redistribute the longwave radiation in the same manner but with higher emissivity $\varepsilon_c$.

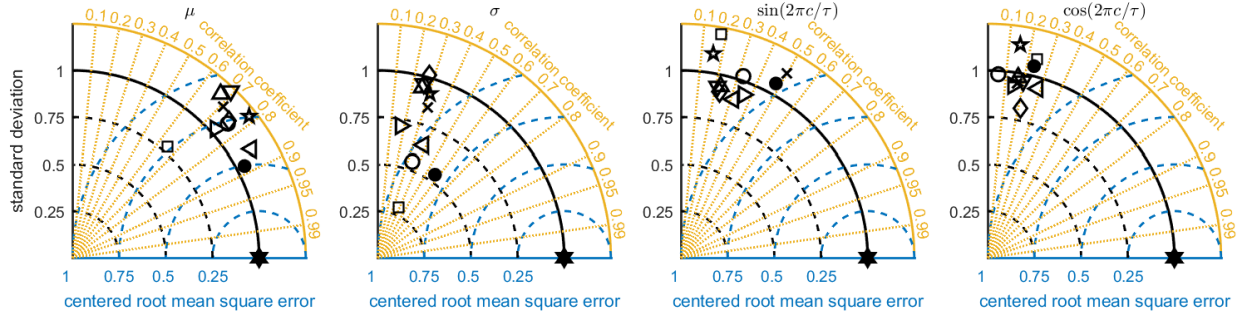

**Fig S4.** Normalized Taylor Diagram for the global spatial patterns of cloud diurnal cycle climatology in winter (December, January, and February). The circular variable centroid (*c*) is converted to cartesian coordinates [$\sin(2\pi/\tau)$, $\cos(2\pi/\tau)$] to produce the corresponding Taylor diagrams. Similar methods have been used for comparing wind field, which is usually decomposed into zonal and meridional components (*15*). For mean ($\mu$) and standard deviation ($\sigma$), the whole global regions with equal-area grids are used for producing the Taylor diagrams; For centroid (*c*), only regions with relative stronger diurnal cycles (here, we assume $c_v$ larger than its 25th percentile) are used for producing the Taylor diagram. Detailed symbols and their corresponding data sources are listed in Table S1.

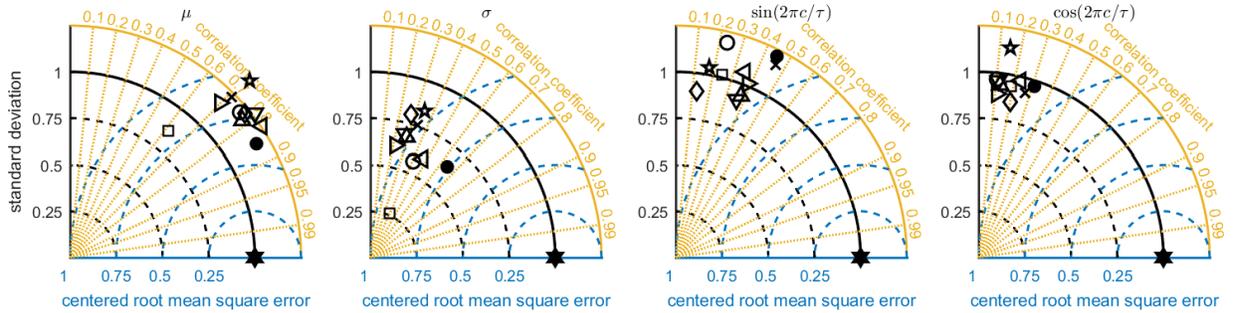

**Fig S5.** As in Fig S4, but for summer (June, July, and August).

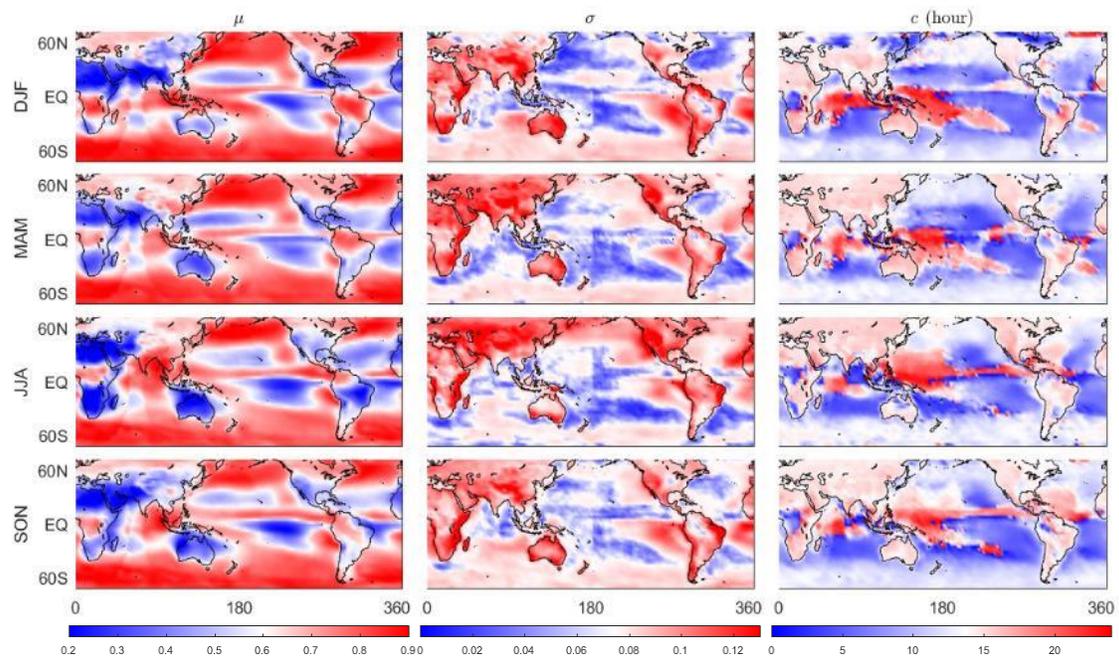

**Fig S6.** Indexes of diurnal cycle of clouds from ISCCP cloud climatology during 1986-2005. Left to right columns show the mean, standard deviation, and centroid of the diurnal cycle of clouds; Top to bottom rows are in winter, spring, summer, and fall seasons.

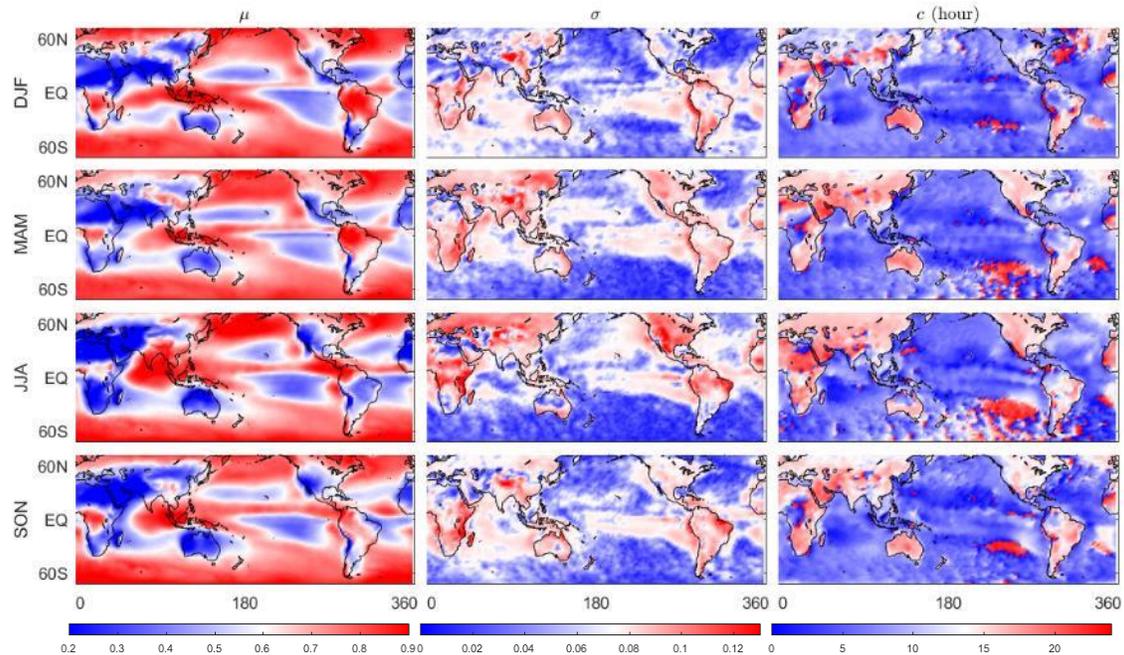

**Fig S7.** As in Fig S6 but for ERA-20C cloud climatology.

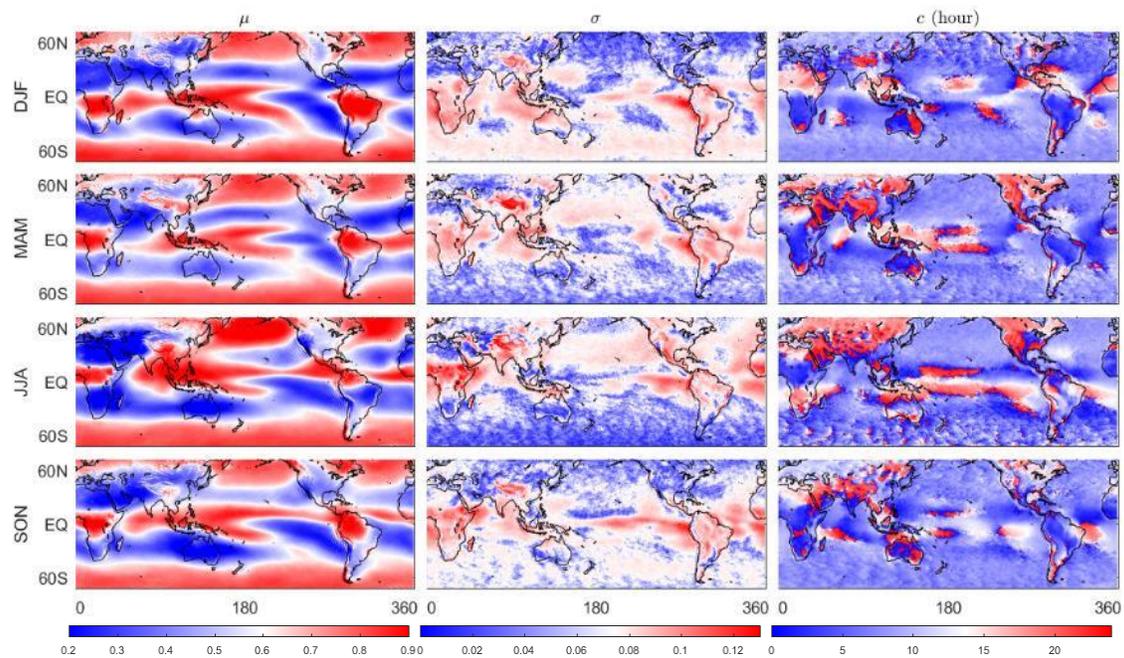

**Fig S8.** As in Fig S6 but for CMCC-CM cloud climatology.

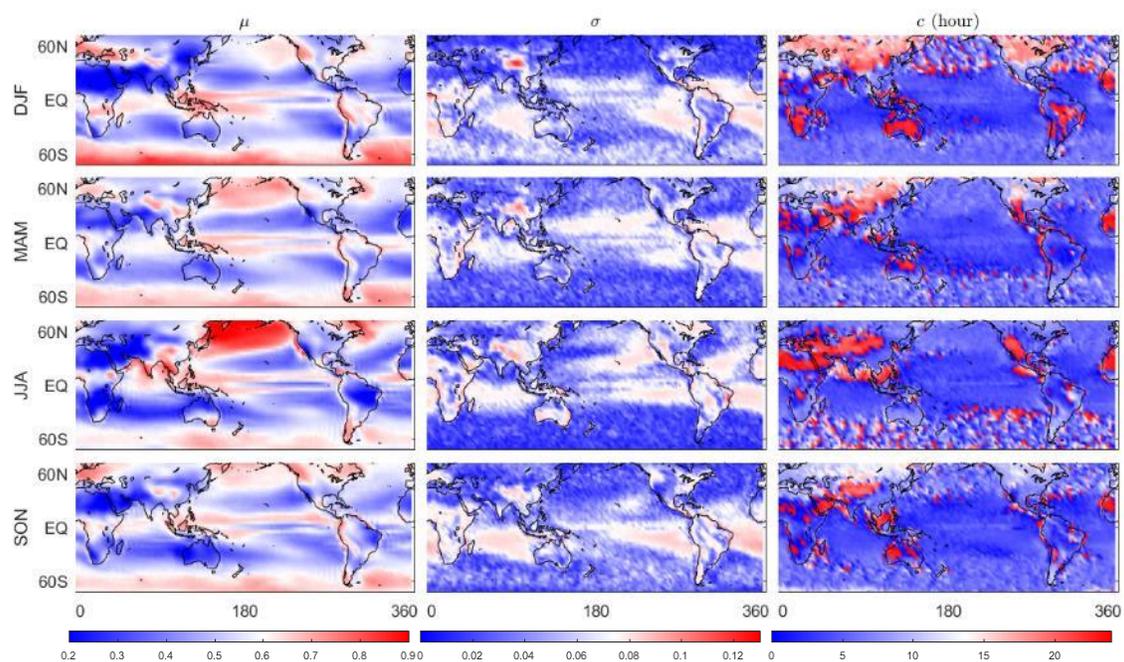

**Fig S9.** As in Fig S6 but for FGOALS-g2 cloud climatology.

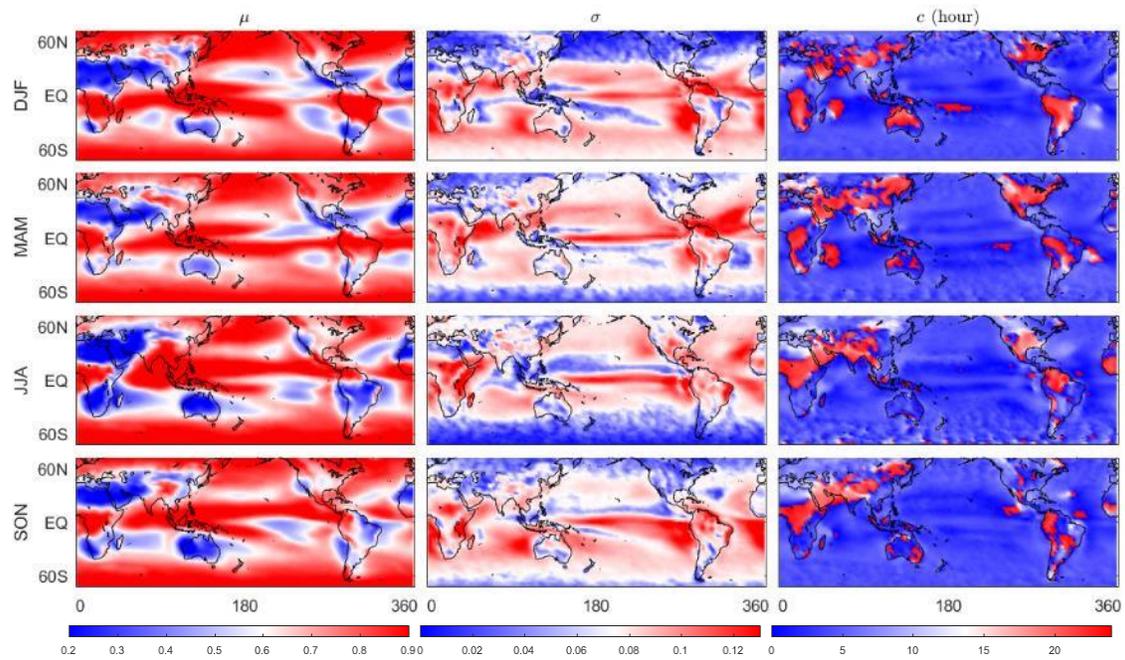

**Fig S10.** As in Fig S6 but for GFDL-CM3 cloud climatology.

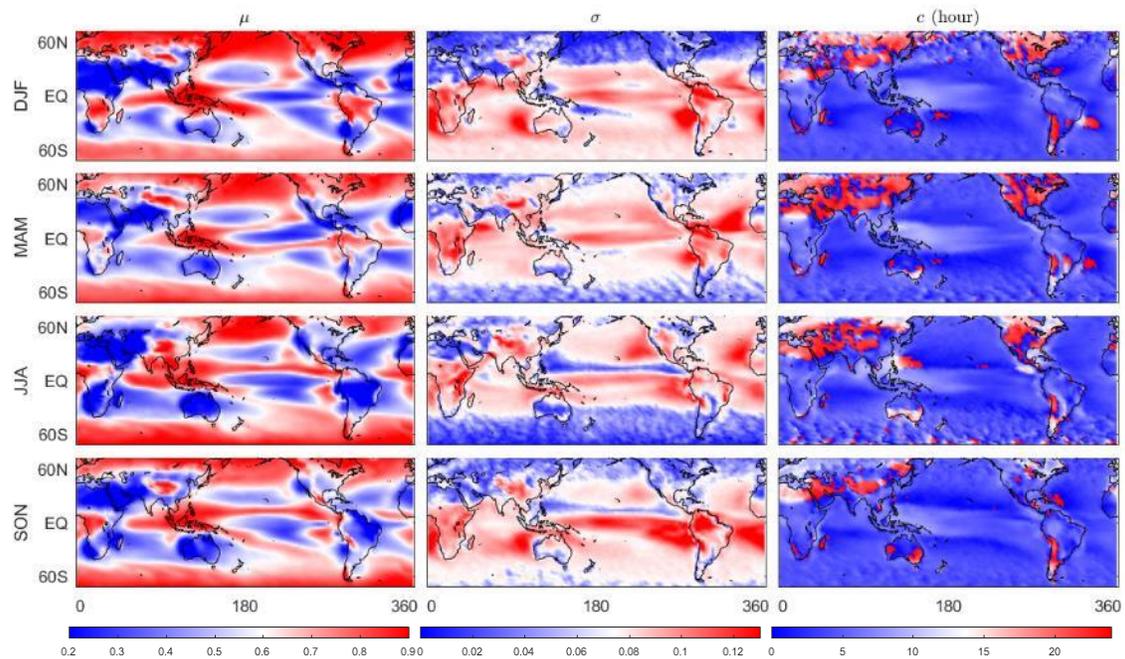

**Fig S11.** As in Fig S6 but for GFDL-ESM2G cloud climatology.

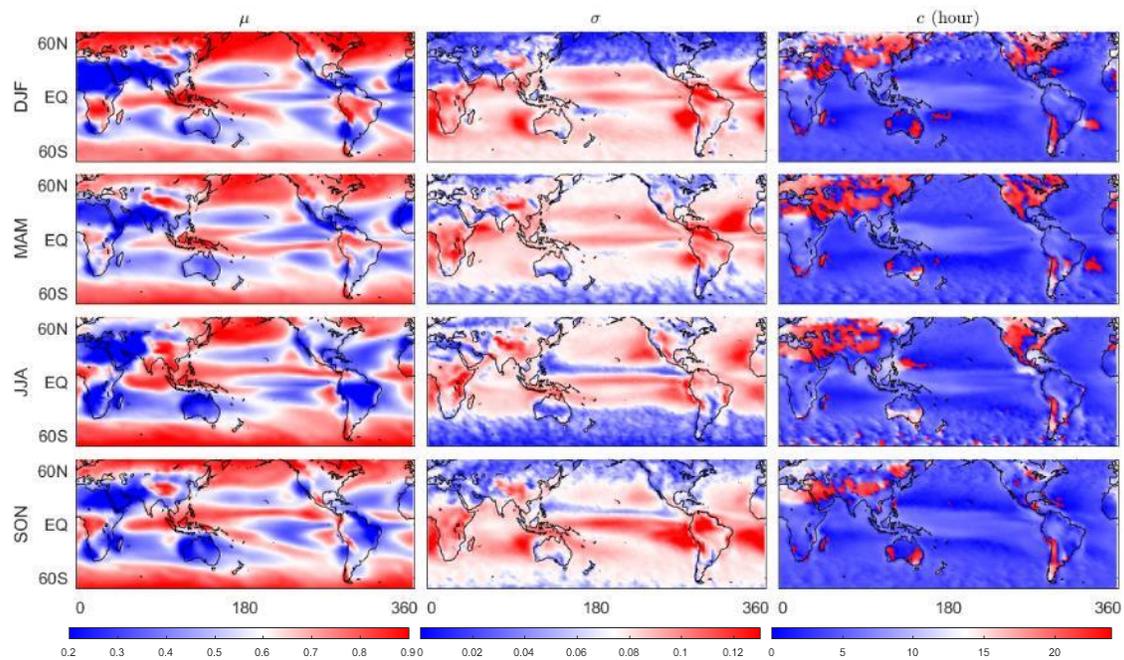

**Fig S12.** As in Fig S6 but for GFDL-ESM2M cloud climatology.

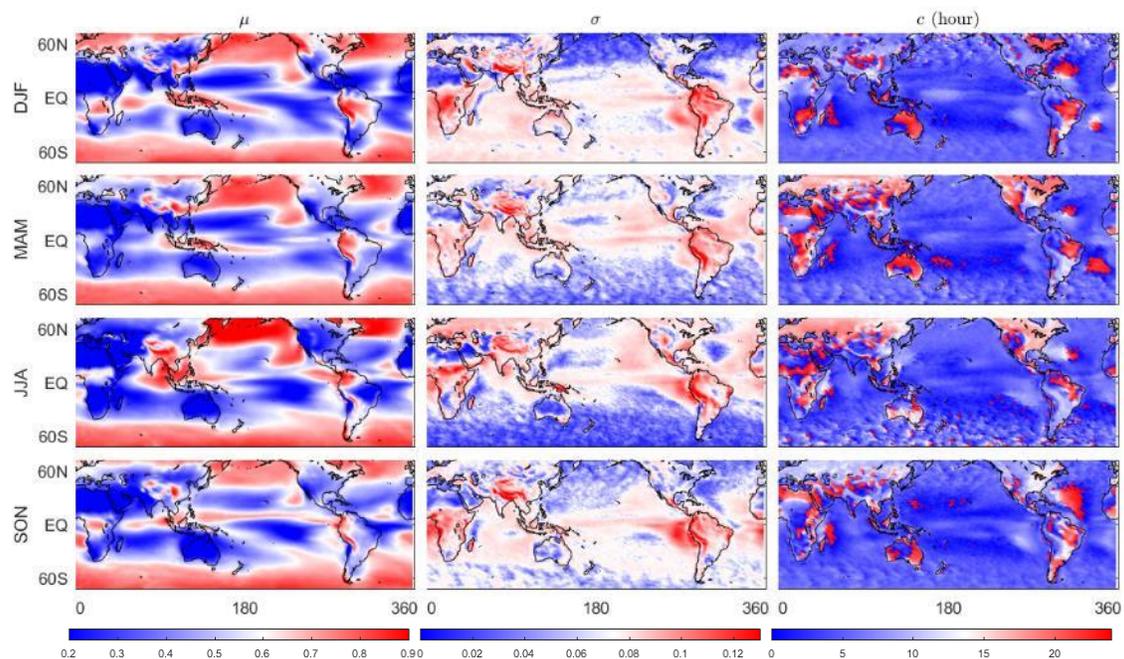

**Fig S13.** As in Fig S6 but for HadGEM2-ES cloud climatology.

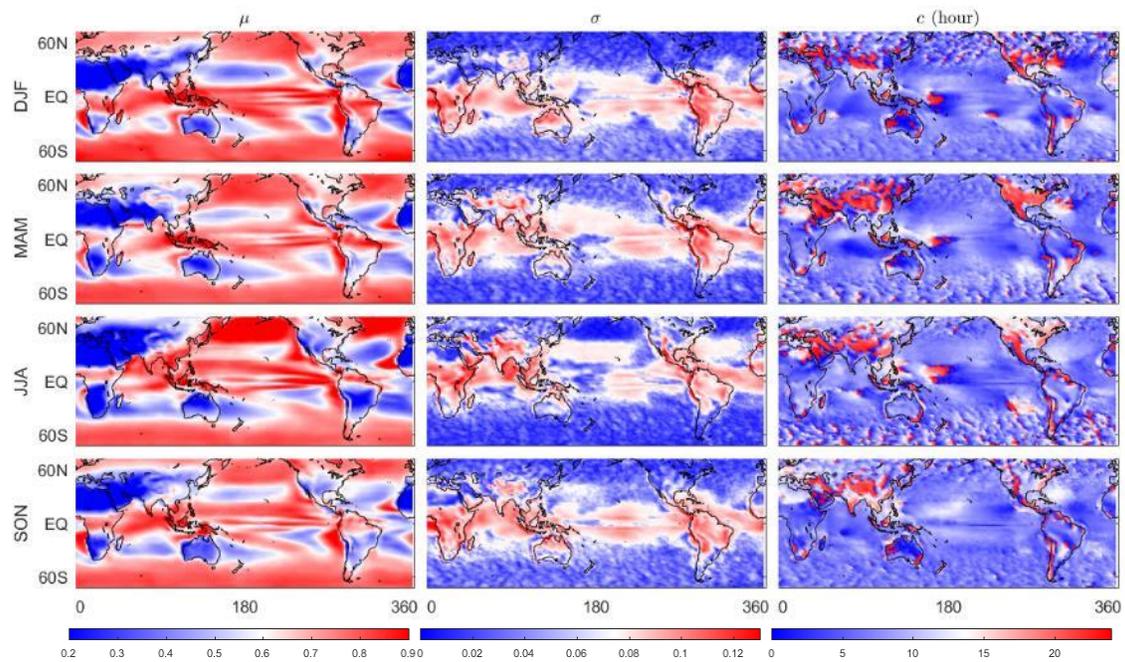

**Fig S14.** As in Fig S6 but for INM-CM4 cloud climatology.

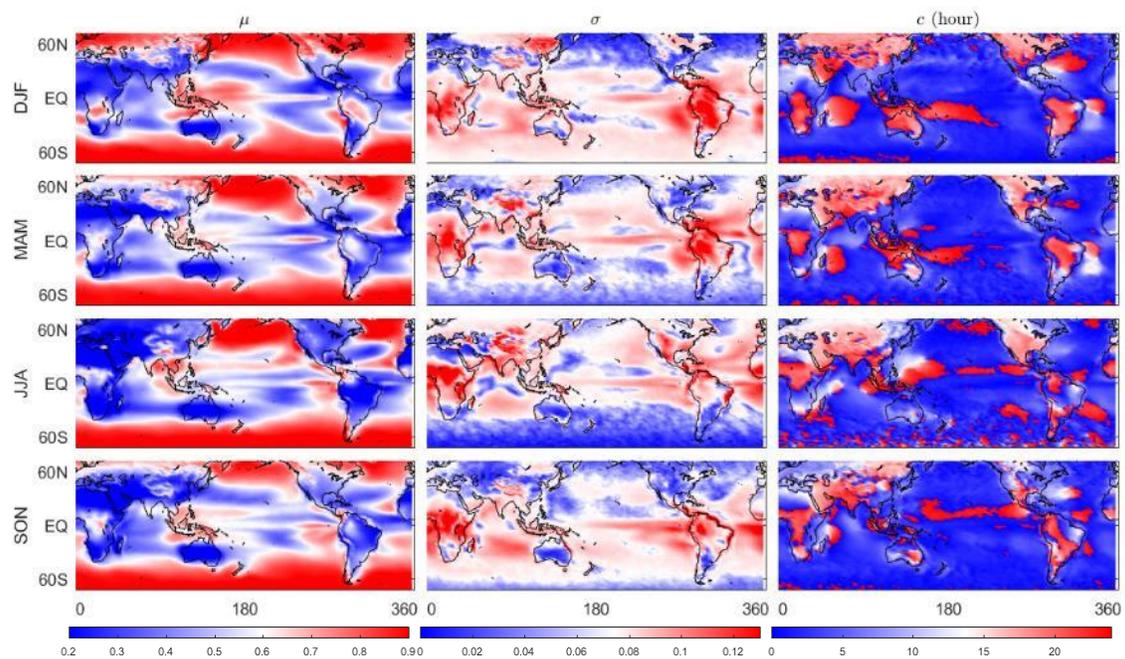

**Fig S15.** As in Fig S6 but for IPSL-CM5A cloud climatology.

**Table S1** Climate Models, satellite observations, and reanalysis data used for assessing cloud diurnal cycles

| # | symbol | acronyms | model institutions and references |
|---|---|---|---|
| 1 | ○ | CMCC-CM | Euro-Mediterranean Center on Climate Change, Italy (*27*) |
| 2 | × | CNRM-CM5 | National Center for Meteorological Research, France (*14*) |
| 3 | □ | FGOALS-g2 | LASG, Institute of Atmospheric Physics, Chinese Academy of Sciences, China; CESS, Tsinghua University, China (*28*) |
| 4 | ◇ | GFDL-CM3 | NOAA Geophysical Fluid Dynamics Laboratory, USA (*29*) |
| 5 | ▽ | GFDL-ESM2G | NOAA Geophysical Fluid Dynamics Laboratory, USA (*30*) |
| 6 | △ | GFDL-ESM2M | NOAA Geophysical Fluid Dynamics Laboratory, USA (*30*) |
| 7 | ◁ | HadGEM2-ES | Met Office Hadley Centre, United Kingdom (*31*) |
| 8 | ▷ | INM-CM4 | Institute for Numerical Mathematics, Russia (*32*) |
| 9 | ☆ | IPSL-CM5A | Institute Pierre Simon Laplace, France (*33*) |
| O | ★ | ISCCP | National Aeronautics and Space Administration, USA (*10*) |
| R | ● | ERA-20C | European Centre for Medium-Range Weather Forecasts (*11*) |